\documentclass[journal]{IEEEtran}
\usepackage{cite}
\usepackage{times}
\usepackage{amsmath,amssymb,amsfonts}
\usepackage{algorithmic}
\usepackage{graphicx}
\usepackage{textcomp}
\usepackage{epstopdf}
\usepackage{amsmath} 
\usepackage{float}
\usepackage{booktabs} 
\usepackage{multirow}   
\usepackage{array}  
\usepackage{graphicx}   
\usepackage{hyperref}
\usepackage{ulem}
\hypersetup{
	colorlinks=true,
	filecolor=blue,      
	urlcolor=blue,
}
\hypersetup{hypertex=true,
	colorlinks=true,
	linkcolor=blue,
	anchorcolor=blue,
	citecolor=blue}
\def\BibTeX{{\rm B\kern-.05em{\sc i\kern-.025em b}\kern-.08em
		T\kern-.1667em\lower.7ex\hbox{E}\kern-.125emX}}

\begin{document}
	
	\title{JOINEDTrans: Prior Guided Multi-task Transformer for Joint Optic Disc/Cup Segmentation and Fovea Detection}
	\author{Huaqing He, Li Lin, Zhiyuan Cai, Pujin Cheng, Xiaoying Tang
		\vspace{-4mm}
		\thanks{This study was supported by the Shenzhen Basic Research Program (JCYJ20190809120205578); the National Natural Science Foundation of China (62071210); the Shenzhen Science and Technology Program (RCYX20210609103056042); the Shenzhen Basic Research Program (JCYJ20200925153847004); the Shenzhen Science and Technology Innovation Committee (KCXFZ2020122117340001). }
		\thanks{Huaqing He, Li Lin, Zhiyuan Cai, Pujin Cheng, Xiaoying Tang are with the Department of Electronic and Electrical Engineering, Southern University of Science and Technology, Shenzhen, China and Jiaxing Research Institute, Southern University of Science and Technology, Jiaxing, China. Corresponding author: Xiaoying Tang (tangxy@sustech.edu.cn).}
	}
	
	\maketitle
	
	\begin{abstract}
		Deep learning-based image segmentation and detection models have largely improved the efficiency of analyzing retinal landmarks such as optic disc (OD), optic cup (OC), and fovea. However, factors including ophthalmic disease-related lesions and low image quality issues may severely complicate automatic OD/OC segmentation and fovea detection. Most existing works treat the identification of each landmark as a single task, and take into account no prior information. To address these issues, we propose a prior guided multi-task transformer framework for joint OD/OC segmentation and fovea detection, named JOINEDTrans. JOINEDTrans effectively combines various spatial features of the fundus images, relieving the structural distortions induced by lesions and other imaging issues.  It contains a segmentation branch and a detection branch. To be noted, we employ an encoder pretrained in a vessel segmentation task to effectively exploit the positional relationship among vessel, OD/OC, and fovea, successfully incorporating spatial prior into the proposed JOINEDTrans framework. There are a coarse stage and a fine stage in JOINEDTrans. In the coarse stage, OD/OC coarse segmentation and fovea heatmap localization are obtained through a joint segmentation and detection module. In the fine stage, we crop regions of interest for subsequent refinement and use predictions obtained in the coarse stage to provide additional information for better performance and faster convergence. Experimental results demonstrate that JOINEDTrans outperforms existing state-of-the-art methods on the publicly available GAMMA, REFUGE, and PALM fundus image datasets.
		We make our code available at \href{https://github.com/HuaqingHe/JOINEDTrans}{https://github.com/HuaqingHe/JOINEDTrans}.

	\end{abstract}
	
	
	\section{Introduction}
	\label{sec:introduction}
\begin{figure}[]
		\centerline{\includegraphics[width=\columnwidth]{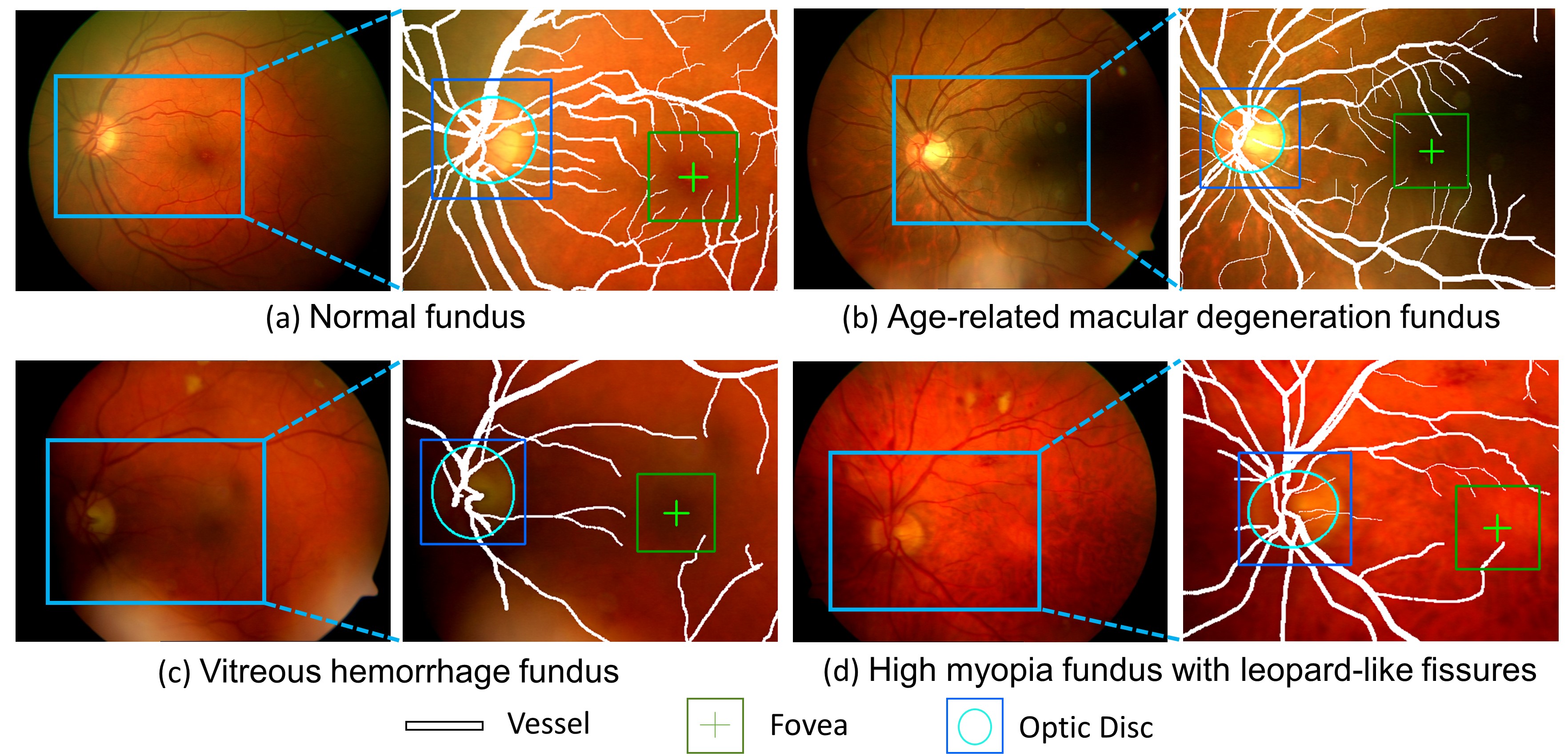}}
		\caption{Consistent spatial relationship among vessel, OD, and fovea in fundus images. (a) Normal fundus with clear OD and fovea; (b) Age-related macula degeneration fundus with obscure macula; (c) Vitreous hemorrhage fundus with obscure OD; (d) High myopia fundus with leopard-like fissures and disappeared macula.}
		\label{fig:fig1}
	\end{figure}

	\IEEEPARstart{R}{etinal} fundus images have been widely used in the clinical diagnoses of various ophthalmic diseases such as glaucoma, age-related macula degeneration, and so on. The optic disc (OD), optic cup (OC), and fovea are critical anatomical landmarks in fundus images and provide important biomarkers of various ophthalmic diseases. For example, the vertical cup-to-disc ratio ($vCDR$) is a key indicator for the diagnosis of glaucoma, and the morphology of fovea is one of the most crucial biomarkers of age-related macula degeneration  \cite{R1,R2}. Therefore, accurate OD/OC segmentation and fovea detection are of great significance for the diagnosis and prognosis of various ophthalmic diseases \cite{R3,R4}. The retinal vessel is another anatimical landmark, which tightly correlates with OD/OC and fovea \cite{R5,R6}. Specifically, the OD/OC region is a hub where vessels concentrate. And the macula whose center is the fovea is an avascular region where vessels end and converge \cite{R44,R45,R46}. 
	Such structural and spatial relationship is determined physiologically and will not be affected by either lesions or image quality issues, as clearly shown in Fig. \ref{fig:fig1}. 
	\par Recently, convolutional neural networks (CNNs) and transformers have been employed for OD/OC segmentation and fovea detection \cite{R8, R9}. 
	Some works focus on designing advanced neural network architectures with powerful semantic extraction capabilities \cite{R10,R11,R12}. They typically treat OD/OC segmentation and fovea detection as separate tasks, and fail to incorporate any prior information. 
	Multi-task learning is naturally employed to accommodate this problem by combining the two tasks of  OD/OC segmentation and fovea detection to make joint predictions \cite{R7,R35}. 
	More recently, research attention has been paid to incorporating various prior information into the OD/OC segmentation as well as the fovea detection tasks. Representative examples include employing the relative positions of OD/OC and fovea in a fundus image to improve the fovea localization accuracy \cite{R13}, utilizing the prior knowledge that vessels and fovea are spatially correlated to enhance the robustness of a fovea localization model \cite{R6}, and so on. However, these works typically design very complex network architectures to incorporate the spatial relationship among fovea, OD/OC, and vessel, at the costs of expensive model storing and training as well as high over-fitting risk. 

	\par To better incorporate structural and spatial prior information from different retinal landmarks in a fundus image, we propose a multi-task learning framework for OD/OC segmentation and fovea detection with an encoder pretrained on vessel segmentation, named JOINEDTrans. On the one hand, multi-task learning can improve the training efficiency. By simultaneously co-optimizing different target functions, the model of interest can learn multi-dimensional representation with relatively few samples \cite{R47}. On the other hand, for different anatomical landmarks in the same fundus image, multi-task learning can fuse location and topology priors through feature or parameter sharing \cite{R14}. 
	We set up a heatmap branch for OD and fovea detection and a segmentation branch for OD/OC segmentation. Through encoder sharing in a multi-task learning setting, the latent representation extracted from the encoder is transmitted to two different decoders. Multi-task learning with a shared encoder can effectively extract implicit associations between OD/OC and fovea, making the model more robust and perform better on multiple tasks \cite{R15, R16}. Meanwhile, considering the structural and spatial relationship between vessels and our target structures of interest (OD/OC and fovea), we conjecture that employing a vessel segmentation pretrained encoder shall well improve the performance and convergence speed of our OD/OC segmentation and fovea detection model. Specifically, we obtain an encoder for vessel segmentation from a data manipulation-based domain-generalized framework \cite{R41}, which guarantees the validity of the vessel features extracted on the target domain in terms of providing effective guidance for localizing OD/OC and fovea.
	\par JOINEDTrans contains a coarse stage and a fine stage. The coarse stage focuses on extracting global information from fundus images, wherein a vessel pretrained Joint Segmentation and Detection Module (JSDM) is designed to obtain coarse OD/OC segmentation and fovea location. The fine stage focuses more on local information, and two modules are respectively proposed for OD/OC segmentation and fovea localization, namely a fine segmentation module (FSM) and a fine localization module (FLM). 
	\par The contributions of this work are summarized as follows:
	\par
	\begin{itemize}
		\item To yield robust outputs in the coarse stage, we propose a multi-task transformer for joint OD/OC segmentation and fovea detection to better extract global information from fundus images. In addition, a vessel segmentation pretrained encoder is used to make better use of prior knowledge.
		\item At the fine stage, we design a multi-branch fovea localization module based on the coarse stage's outputs through coordinate regression and heatmap localization. Meanwhile, to refine the OD/OC segmentation and fovea localization results, we take cropped OD/OC segmentation and fovea heatmap from the coarse stage as additional inputs into the fine stage, greatly boosting the efficiency and accuracy of our two tasks of interest.
		\item We extensively evaluate JOINEDTrans on three publicly available fundus image datasets. Experimental results show that our method achieves state-of-the-art (SOTA) performance on both OD/OC segmentation and fovea detection. Notably, JOINEDTrans achieves significant improvements with much fewer parameters compared to other transformer-based methods trained from scratch.
		
	\end{itemize}
	This paper extends our previous conference work JOINED \cite{R13} in several aspects:
	
	\begin{itemize}
		\item Incorporating richer prior information, JOINEDTrans significantly improves over JOINED. Most of the multi-scale features learned by the original JOINED framework are related to the spatial positions of OD and fovea, and thus are highly correlated with each other.
		In this work, we remove the branch that predicts the distance of each pixel to OD or fovea. We innovatively incorporate the positional relationship among vessel, OD, and fovea through a pretrained encoder, which largely simplifies the entire structure of the model and benefits an extraction of more complementary positional information.
		\item We replace CNN with transformer given that transformer can better identify global representation when fusing decoder features. Such a replacement induces large improvements in the coarse stage; the coarse stage results from JOINEDTrans even approach the fine stage results from JOINED. 
		\item Since both OD and OC are elliptical-like structures, we novelly introduce a star-shape loss when training the network, which speeds up segmentation convergence  and corrects erroneous segmentation results.
		\item Finally, we conduct more comprehensive experiments to verify the effectiveness of the newly proposed JOINEDTrans. Specifically, we compare with more SOTA methods and conduct more detailed ablation analyses.
		
	\end{itemize}
	
	\section{Related Works}
	
	\subsection{Fundus based OD/OC Segmentation and Fovea Detection}
	\par
	Fundus based OD/OC segmentation and fovea detection are crucial steps in diagnosing glaucoma and other related ophthalmic diseases. For OD/OC segmentation, many methods have been proposed. For instance, 
	Fu et al. \cite{R79} employ polar coordinates to perform joint OD/OC segmentation, taking into account their area ratio. 
	Zheng et al. \cite{R25} present a multi-scale CNN to first generate an initial contour and then deliver more accurate shape and boundary details.
	Pachade et al. \cite{R74} design an enhanced patch-based discriminator in Nested EfficientNet to polish the boundaries of OD/OC segmentation. 
	Meng et al. \cite{R31} explore the properties of graph models to expand the receptive area for representing long-range context relationships. These methods generally achieve promising OD/OC segmentation results. 
	\par
	For fovea detection, before CNNs are widely employed for medical image detection, researchers typically utilize handcrafted features equipped with traditional machine learning techniques \cite{R60,R61}. Medhi et al. \cite{R60} and Deka et al. \cite{R61} make use of the vessel information to generate regions of interest (ROIs) for subsequent macula estimation. 
	These traditional methods are relatively less accurate but more robust due to their employing the spatial relationship between vessel and fovea. 
	Some deep learning-based methods also impose such spatial relationships as constraints for fovea localization. For example, Huang et al. \cite{R2} incorporate the positional relationship between OD and fovea into a CNN model to refine and narrow the fovea ROI.
	After identifying vessels and OD through UNet and probability bubbles, Fu et al. \cite{R65} summate all the vessel and OD vectors to obtain the retinal raphe, and then estimate the fovea location through local region searching. Meyer et al. \cite{R57} employ a pixelwise regression method for joint OD and fovea detection, wherein the dependent variable is the distance from each pixel to the nearest interested landmark. 
	Song et al. \cite{R6} propose a transformer-based fovea detection network which incorporates vessels' spatial information to improve robustness. 
	More recently, Sedai et al. \cite{R59} propose a two-stage segmentation framework, which first locates a rough position of the macula, and then precisely locates the fovea.
	Xie et al. \cite{R77} perform multi-scale feature fusion combined with self-attention for end-to-end fovea prediction.
	These methods also attain promising results on fovea detection.
	However, there is still space for improvements in terms of both OD/OC segmentation and fovea detection, especially in cases of lesions or low image qualities.
	
	\subsection{Multi-task Learning for Fundus Segmentation}
	\par
	Multi-task learning for fundus segmentation improves a model's generalizability through parallel learning tasks, which can be roughly divided into two categories. The first category is to extract features by combining different tasks targeting a same structure.
	For example, Meng et al. \cite{R31} use a graph reasoning module to extract features that represent the region-and-boundary relationship for OD/OC. 
	Sun et al. \cite{R32} learn OD/OC's boundary features through multi-scale feature extraction. 
	The other category focuses on multi-task information sharing across different anatomical structures. For example, Ana et al. \cite{R5} make use of the relationship between vessels and OD, and employ traditional reconstruction algorithms to remove vessels in the OD area. Tan et al. \cite{R7} utilize a single network to simultaneously segment vessels, fovea, and OD, and implicitly extract their relationship. Pascal et al. \cite{R35} combine features extracted from OD/OC segmentation and fovea localization to benefit glaucoma classification. 
	Although these methods can well utilize the complementary information between OD/OC and other anatomical structures, they tend to focus on incorporating complex blocks or modifying network structures.
	In this paper, we propose a multi-task framework targeting vessels, fovea, and OD/OC. We employ an encoder pretrained on vessel segmentation \cite{R41} to implicitly incorporate vessel information.
	\subsection{Prior Guided Fundus Segmentation}
	\par
	Prior knowledge can guide a model of interest to extract features purposefully, and thus effectively improve the model's robustness and accuracy. In fundus segmentation, the prior knowledge can be generally categorized into size related, topology related, and shape related \cite{R80}.  
	The first category incorporates the size information of various fundus structures to a segmentation model and limits the predicted structure area to a specific size range \cite{R32, R77}.  
	The second category utilizes topological  priors, such as fundus structures' connectivity patterns \cite{R76,R82}  and the relative position constraints among different structures \cite{R2,R6,R13}. 
	The last category is usually computed on the shape of a structure of interest which is represented by a set of points or a binary map. Most exiting works integrate shape descriptors into CNNs' decoders or loss terms. A representative type of shape prior is the star-shape prior. Liu et al. \cite{R38, R39} employ the star-shape prior on skin cancer detection and fundus segmentation to optimize the output by getting combined, in a simple and straightforward manner, with DeepLabv3+. Camarasa et al. \cite{R37} also use the star-shape prior for weakly supervised segmentation on carotid arteries, which greatly benefits segmentation.
	
	 
	Through incorporating prior information, better performance on fundus segmentation is generally achieved. But most existing works incorporate priors by modifying models' architectures or loss terms, which is relatively complicated. 
	Different from those previous works, we use an encoder pretrained on vessel segmentation to implicitly introduce the vessel information and combine OD/OC's location information for fovea detection. We also employ the star-shape loss as part of our OD/OC segmentation loss function to polish their shapes.
	We demonstrate that our proposed strategy outperforms existing methods.

	\begin{figure*}[!t]
		\centering
		\includegraphics[width=\linewidth]{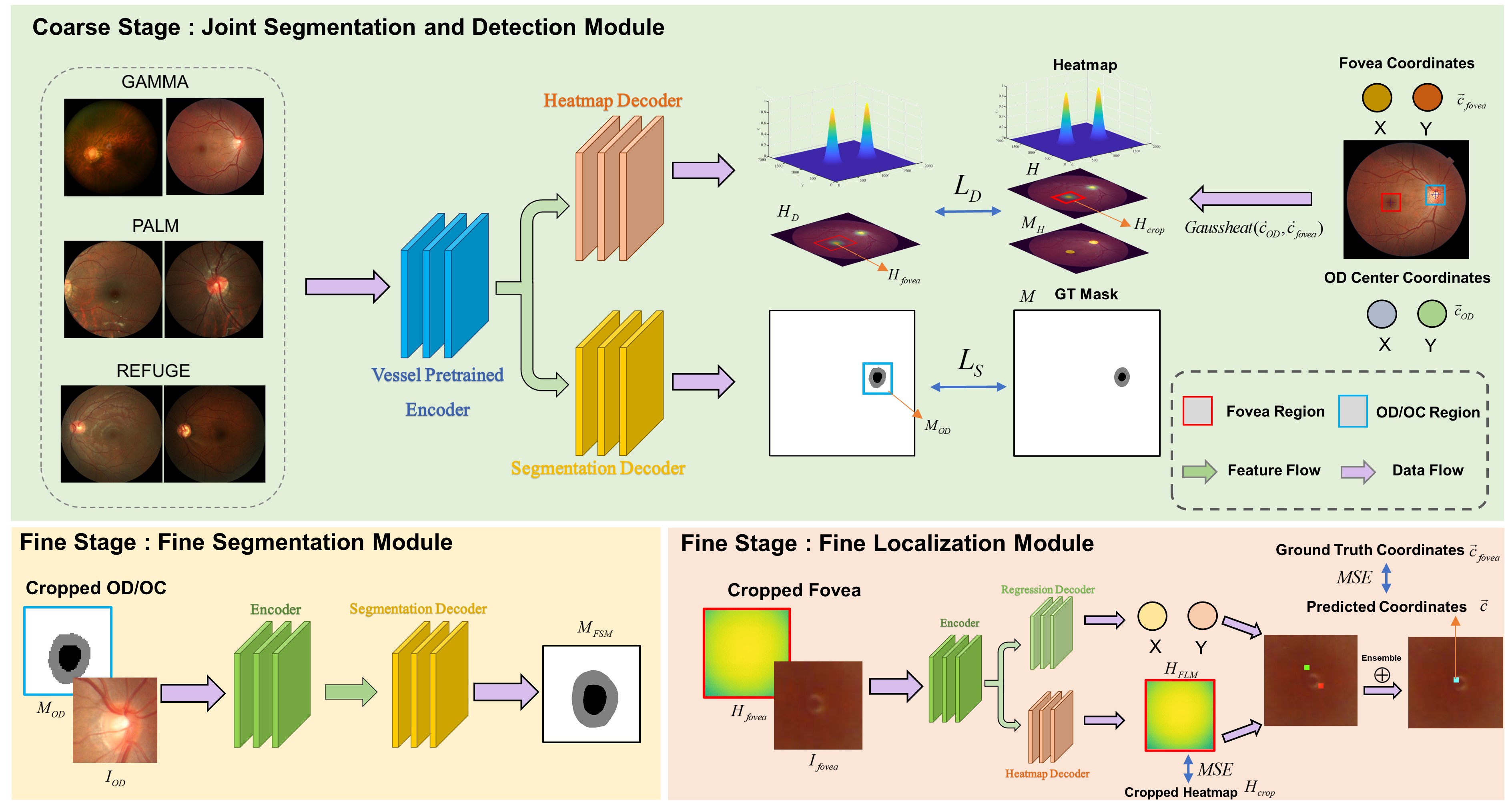}
		\caption{Overview of the proposed JOINEDTrans framework. The vessel pretrained encoder loads pretrained parameters obtained from AADG \cite{R41}. In the coarse stage, the locations and heatmap of OD's center and fovea are obtained through the heatmap decoder, and the coarse segmentation results of OD/OC are obtained through the segmentation decoder. These coarse heatmap and segmentation results are then respectively used as additional inputs into the fine localization and fine segmentation modules in the fine stage.}
		\label{fig:framework}
	\end{figure*}

	\section{Methodology}
	The proposed JOINEDTrans consists of three modules: JSDM, FLM for fovea detection, and FSM for OD/OC segmentation, with two stages: a coarse stage and a fine stage. Fig. \ref{fig:framework} shows its overall pipeline. 
	\subsection{Joint Segmentation and  Detection Module}
	\par 
	
\begin{figure*}[!t]
		\centering
		\includegraphics[width=\linewidth]{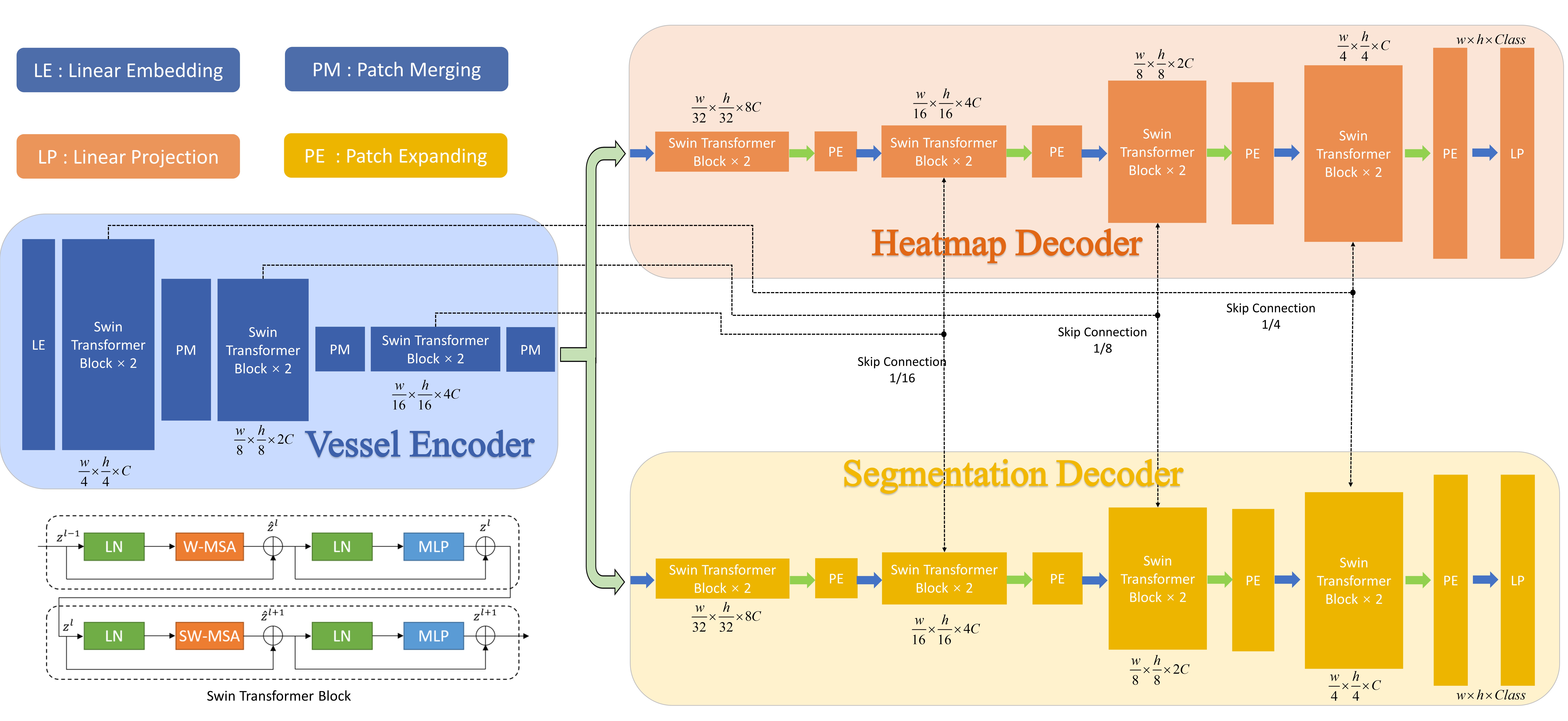}
		\caption{The architecture of JSDM, which is composed of a vessel encoder, a heatmap decoder, a segmentation decoder and skip connections. The encoder and decoders are all based on Swin-Transformer blocks.}
		\label{fig:architecture}
	\end{figure*}

	JSDM is composed of a vessel pretrained encoder, a heatmap detection decoder, a segmentation decoder and skip connections, as shown in Fig. \ref{fig:architecture}.
	An input fundus image is divided into non-overlapping patches. Each patch is treated as a token and inputted into the vessel pretrained encoder to learn deep feature representation.
	The two decoders employ a patch expanding layer to up-sample the extracted context features, upon which skip connection based fusion is performed with the multi-scale features from the encoder to restore the spatial resolution of the feature mapping. After that, segmentation prediction and heatmap reconstruction are conducted.
	\par Assume we are given a training set $T=\{ I^i, M^i , H^i \}^N_{i=1} $, where $I^i$ is a fundus image, $ M^i $	is the corresponding ground truth OD/OC segmentation, and $ H^i $ is a heatmap image constructed with the coordinates of the fovea and the OD center through a Gaussian kernel matrix $ \mathcal{G}(\cdot) $. For OD/OC segmentation, the goal is to estimate their segmentation mask $ M $.
	For fovea detection, the goal is to estimate its coordinates $ C = [X, Y]  $.  
	JSDM contains two decoder branches that share a common encoder, working in a manner as
	\begin{equation}\label{eq:JSDM} 
	[H_D,P_S]=\mathcal{F}_{JSDM}(I;\theta_{JSDM}),
	\end{equation}
	where $ \theta_{JSDM} $ denotes the parameters of $ \mathcal{F}_{JSDM} $,  $H_D$ is the output of the detection branch and $ P_S $ 
	is a probability map generated from the segmentation branch.

	\subsubsection{Vessel Pretrained Encoder}
	In the vessel pretrained encoder, a $C$-dimensional tokenized input with a resolution of $w/4 \times h/4$ is fed into two consecutive Swin-Transformer blocks for representation learning, where $w$ and $h$ respectively represent the weight and height of the image of interest, and $C$ represents the dimension of the tokenized feature. The Swin-Transformer blocks enable representation learning without changing the feature dimension and resolution.
	Meanwhile, the patch merging layer reduces the number of tokens ($2\times$ downsampling) and doubles the feature dimension. This procedure gets repeated three times in the encoder.
	
	\textbf{Patch merging layer :} The input patches are divided into four parts and concatenated by a patch merging layer. Through such processing, the feature resolution is downsampled by 2$\times$. In addition, since the feature dimension increases by $4\times$ due to the concatenation operation, a linear layer is applied to the concatenated features so as to unify the feature dimension to be $2\times$ the original dimension.

	We use the publicly-accessible CHASEDB, DR-HAGIS, DRIVE, HRF, LES-AV, Aria-ACD datasets \cite{R66, R67, R68, R69, R70, R71}, containing a total of 338 samples, for training the vessel encoder.
	We adopt a data-driven vessel segmentation model \cite{R41}, which automatically searches for optimal data augmentation operations for the target domain and has strong generalizability accommodating different domains.
	The wide image domains of the six training datasets ensures that the vessel encoder can extract robust vascular features for fundus images from various unseen image domains. These vascular characteristics shall effectively guide the OD/OC segmentation and fovea localization tasks.
	
	\subsubsection{The HeatMap Detection Branch}
	In the heatmap detection branch, the outputted heatmap $H_D$ has two layers, one representing the heatmap for OD/OC and the other for fovea. 
	The heatmaps are generated by two Gaussian kernel matrices  $\mathcal{G}(\vec{c}_{_{OD}})$ and $\mathcal{G}(\vec{c}_{fovea})$,
	\begin{equation}\label{eq:example}
	\mathcal{G}(\vec{c_k})=\frac{1}{2\pi\sigma^2}e^{-\frac{||\vec{x}-\vec{c_k}||^2_2}{2 \sigma ^2}},  k \in \{OD\; center,fovea\},
	\end{equation}
	where $\sigma$ is set to be $h/20$, with $h$ being the image height. If there exists no OD or fovea due to poor image quality, we  set the value of the corresponding $\vec{c}_{_{OD}}$ or $\vec{c}_{fovea}$ to 0.  
	After obtaining  $\mathcal{G}(\vec{c}_{_{OD}})$ and $\mathcal{G}(\vec{c}_{fovea})$, we normalize them to [0,1] and concatenate them together to form the detection branch's ground truth $ H $. Specifically,
	\begin{equation}\label{eq:example} 
	H=concatenate\big(\mathcal{G}(\vec{c}_{_{OD}}), \mathcal{G}(\vec{c}_{fovea})\big).
	\end{equation}
	 We identify the predicted coordinates of the fovea and the OD center to be the coordinates of the pixels having the maximum heatmap values. If the detection process does not succeed, we approximate fovea through a pre-specified relationship \cite{R2} between fovea and OD center.  
	The objective function $\mathcal{L}_D$ of the detection branch is defined as 
	
	\begin{equation}\label{eq:L_D} 
	\mathcal{L}_D = MSE(H,H_D)+Dice(M_H, H_D),
	\end{equation}
	where $H$ is the ground truth heatmap of fovea and OD center.
	$M_H$ is the binary form of $H$ through thresholding at 0.5, which corresponds to the heatmap's ground truth segmentation of the fovea and OD region. $MSE$ is employed as the heatmap prediction loss and $Dice$ is employed as the heatmap segmentation loss. These two losses together form the final heatmap detection loss.
	\par 
	\textbf{Patch expanding layer :} Before up-sampling, we apply a linear layer to the input features ($w/32 \times h/32 \times 8C$) to double the feature dimension ($w/32 \times h/32 \times 16C$). Then,  we employ  a rearrangement operation to twice the input resolution and reduce the feature dimension to a quarter of the input dimension ($w/32 \times h/32 \times 16C \to w/16 \times h/16 \times 4C$).
	\subsubsection{The Segmentation Branch} The segmentor outputs a probability map $P_S\in [0,1]^{h \times w \times 3}$, where the three layers respectively represent the probabilities of OC, OD, and the background. To obtain OD/OC's coarse segmentation result $M_S$, we set the threshold in all layers of $P_S$ to 0.5.
	Since it is a multi-classification problem, some pixels may be simultaneously classified as both OD and OC. Under such circumstances, we set the priority of category classification as OC $>$ OD, since OD always spatially contains OC. We employ $Dice$ to be the segmentation branch's loss. 

	%
	
	\par In the training phase, the heatmap detection branch is utilized to detect the rough locations of OD and fovea to help more stable OD/OC segmentation. Therefore, the final loss of JSDM is defined as (with coefficients $\lambda_0$, $\lambda_1$ used to balance the two terms) 
	\begin{equation}\label{eq:L_JSD}
	\mathcal{L}_{JSDM} = \lambda_0\mathcal{L}_D + \lambda_1\mathcal{L}_S,
	\end{equation}
	where $\mathcal{L}_S$ is explained in Eq. \ref{eq:L_Star}.
	\subsubsection{Skip Connection} Similar to UNet, skip connections are used to fuse the multi-scale features from the encoder with the up-sampled features. We concatenate the shallow features and the deep features together to reduce the loss of spatial information caused by down-sampling. The overall setting is consistent with that in Swin-Unet \cite{R18}, wherein the dimension of the connected features is the same as that of the upsampled features. 

	\subsection{Star-Shape Loss} 
	To further enhance OD/OC's segmentation quality, we make use of the prior knowledge that both OD and OC are star-shaped, or more specifically elliptical. The star-shape loss encourages continuous segmentation with smooth boundaries, which well accommodates OD/OC's prior of being elliptical. 
	\par We make use of such prior information and impose additional regularization, in the form of a star-shape loss, to normalize the output of the segmentation branch.
	Specifically, for each potential segmentation result, we extract its boundary and centroid. After that, we calculate and minimize the variance of the boundary-to-centroid distances. 
	The smaller the variance, the more convex and continuous the segmentation. 
	To back-propagate model parameters, we employ angle as a differentiable way to represent points on the boundary, ending up with a star-shape loss. 
	We refer readers to \cite{R37} for more details on the star-shape loss. The loss of the coarse stage's segmentation branch thus becomes
	\begin{equation}\label{eq:L_Star} 
	\mathcal{L}_S=\mu Dice(M, P_S)+\nu \mathcal{L}_{star}, 
	\end{equation}
	with $\mu$ and $\nu$ being hyper-parameters balancing the $Dice$ loss and the star-shape loss $\mathcal{L}_{star}$, and  $M$ is OD/OC's ground truth segmentation.
	\subsection{Fine Segmentation Module} FSM utilizes an adapted Swin-Unet \cite{R18} pretrained on ImageNet \cite{R50} as the segmentation network, which produces OD/OC's final segmentation result $ M_{FSM} $. Specifically $[M_{FSM}] = \mathcal{F}_{FSM}(I_{OD},M_{OD};\theta_{FSM})$, where $M_{OD}$ is a cropped result of the coarse OD/OC segmentation obtained from the JSDM segmentation branch and $I_{OD}$ is the corresponding cropped fundus image masked by  $M_{OD}$. 
	We input not only $I_{OD}$ into FSM, but also $M_{OD}$ for faster convergence and better accuracy.  
	The loss function of FSM is the same as that in the coarse stage's segmentation branch. 
	
	\subsection{Fine Localization Module} A multi-task learning strategy is adopted in FLM. Specifically $ [\vec{c}, H_{FLM}] = \mathcal{F}_{FLM}(I_{fovea},H_{fovea};\theta_{FLM})$, where $H_{FLM}$ is the predicted heatmap of the cropped fovea region. 
	$\vec{c}$ is the predicted coordinates of fovea. We employ FLM to produce predicted fovea coordinates as well as accurate estimates of $H_{FLM}$ simultaneously for all pixels. Comparing the results of the two branches, if the Euclidean distance between the two exceeds a threshold $d_{FLM}$, the regression result is adopted. Otherwise, the mean of the heatmap and the regression results is adopted to get the final coordinates $\vec{c}$. 
	$H_{fovea}$ is the fovea region of $H_D$ from the coarse stage. We concatenate $H_{fovea}$ with the corresponding cropped  fundus image $I_{fovea}$ as the input.

	\par FLM's loss function is defined in Eq. \ref{eq:L_FLM} and $\vec{c}_{fovea}$ is the ground truth of fovea's coordinates. The finally predicted coordinates of fovea are an ensemble of $\vec{c}$ and the coordinates obtained from $H_{FLM}$. $H_{crop} $ is the corresponding cropped version of $H$.
	\begin{equation}\label{eq:L_FLM}
	\begin{aligned}
	\mathcal{L}_{FLM} & = \mathcal{L}_{regression} + \mathcal{L}_{heatmap}\\
	& = MSE(\vec{c}_{fovea},\vec{c})+MSE(H_{crop},H_{FLM}).
	\end{aligned}
	\end{equation}

	\begin{table*}[t]
		\centering
		\caption{Performance comparisons between our proposed JOINEDTrans and other SOTA methods, as evaluated on the GAMMA dataset. The best results are highlighted in bold, and the second-best results are underlined.}
		\label{tab:results-gamma}
		\resizebox{\textwidth}{!}{
		\begin{tabular}{lcccccccccccc}
			
			\specialrule{0.5mm}{1pt}{1pt}
			\multicolumn{13}{c}{GAMMA}                                                                                                                                                                                                                             \\  \specialrule{0.05mm}{1pt}{1pt}
			\multicolumn{1}{c}{\multirow{3}{*}{Method}} & \multicolumn{6}{c}{Coarse}                                                                                             &  & \multicolumn{5}{c}{Fine}                                                                         \\ \cline{2-7} \cline{9-13} 
			\specialrule{0em}{1pt}{1pt}
			\multicolumn{1}{c}{} & \multicolumn{2}{c}{Detection}               &  & \multicolumn{3}{c}{Segmentation}                                      &  & Detection             &  & \multicolumn{3}{c}{Segmentation}                                      \\  
			\multicolumn{1}{c}{} & Fovea $AED$ $\downarrow$ & OD $AED$ $\downarrow$ &  & OD $Dice$ (\%) $\uparrow$ & OC $Dice$ (\%) $\uparrow$ & $vCDR$ (\%) $\downarrow$ &  & Fovea $AED$ $\downarrow$ &  & OD $Dice$ (\%) $\uparrow$ & OC $Dice$ (\%) $\uparrow$ & $vCDR$ (\%) $\downarrow$ \\ \specialrule{0.05mm}{1pt}{1pt}
			DeepLabv3+  \cite{R40}           & -                     & -                   &  & 78.27$\pm$18.34             & 70.41$\pm$20.27             & 17.22$\pm$10.25       &  & -                     &  & 86.22$\pm$12.25             & 72.80$\pm$26.15             & 15.24$\pm$10.81       \\
			UNet++  \cite{R29}              & -                     & -                   &  & 77.22$\pm$16.22             & 73.29$\pm$21.18             & 16.31$\pm$11.18       &  & -                     &  & 85.41$\pm$15.27             & 75.09$\pm$24.25             & 16.93$\pm$15.22       \\
			UNet  \cite{R30}                & -                     & -                   &  & 80.22$\pm$10.87             & 74.75$\pm$22.28             & 15.22$\pm$10.27       &  & -                     &  & 87.15$\pm$10.87             & 76.54$\pm$23.90             & 14.21$\pm$11.24       \\
			UNet (ResNet50)         & 46.67$\pm$146.28          & 31.93$\pm$171.59        &  & 79.19$\pm$17.92             & 72.54$\pm$20.00             & 16.71$\pm$11.36       &  & -                     &  & 88.45$\pm$9.38              & 75.46$\pm$14.69             & 15.45$\pm$11.39       \\
			UNet++ (ResNet50)       & 43.15$\pm$162.56          & 32.93$\pm$178.84        &  & 81.71$\pm$15.91             & 70.21$\pm$19.14             & 70.21$\pm$19.14       &  & -                     &  & 89.94$\pm$6.58              & 73.24$\pm$21.43             & 14.43$\pm$10.24       \\
			DeepLabv3+ (ResNet101)  & 39.27$\pm$84.14           & 35.14$\pm$48.52         &  & 82.17$\pm$13.52             & 74.91$\pm$16.30             & 12.71$\pm$10.00       &  & -                     &  & 91.28$\pm$7.38              & 79.46$\pm$9.87              & 12.75$\pm$9.47        \\ \specialrule{0.05mm}{1pt}{1pt}
			Pixel-Wise Regression \cite{R57}   & 35.16$\pm$53.15           & 40.71$\pm$30.87         &  & -                       & -                       & -                 &  & 20.16$\pm$25.19           &  & -                       & -                       & -                 \\
			H-DenseUNet \cite{R56}             & -                     & -                   &  & 80.51$\pm$11.95             & 78.46$\pm$15.25             & 11.48$\pm$10.73       &  & -                     &  & 93.57$\pm$6.61              & 84.25$\pm$7.99              & 9.242$\pm$7.22        \\
			Attention UNet \cite{R10}       & -                     & -                   &  & 83.53$\pm$15.49             & 75.27$\pm$15.02             & 13.22$\pm$11.27       &  & -                     &  & 94.22$\pm$7.10              & 85.30$\pm$5.62              & 7.412$\pm$6.19        \\
			JOINED  \cite{R13}                  & \renewcommand{\ULdepth}{0.7pt}\uline{31.15$\pm$54.27}           & \renewcommand{\ULdepth}{0.7pt}\uline{32.93$\pm$28.84}         &  & \renewcommand{\ULdepth}{0.7pt}\uline{92.85$\pm$5.28}              & \renewcommand{\ULdepth}{0.7pt}\uline{82.86$\pm$9.49}              & 9.436$\pm$6.40        &  & \renewcommand{\ULdepth}{0.7pt}\uline{15.15$\pm$30.56}           &  & \renewcommand{\ULdepth}{0.7pt}\uline{95.53$\pm$5.60}              & 86.89$\pm$9.10              & \renewcommand{\ULdepth}{0.7pt}\uline{3.938$\pm$2.24}        \\ \specialrule{0.05mm}{1pt}{1pt}
			Segtran  \cite{R28}             & -                     & -                   &  & 83.12$\pm$16.27             & 75.24$\pm$22.70             & 9.150$\pm$8.51        &  & -                     &  & 88.78$\pm$15.06             & 79.41$\pm$18.33             & 9.982$\pm$8.18        \\
			TransUnet  \cite{R19}               & -                     & -                   &  & 90.10$\pm$8.32              & 79.98$\pm$13.53             & \renewcommand{\ULdepth}{0.7pt}\uline{7.926$\pm$7.19}        &  & -                     &  & 94.72$\pm$4.31              & 85.97$\pm$10.09             & 5.926$\pm$4.92        \\
			Swin-Unet  \cite{R18}                & -                     & -                   &  & 84.91$\pm$13.03             & 73.88$\pm$20.48             & 10.98$\pm$9.71        &  & -                     &  & 94.97$\pm$6.24              & \renewcommand{\ULdepth}{0.7pt}\uline{88.85$\pm$8.93}              & 3.986$\pm$3.93        \\
			Proposed                & \textbf{25.24$\pm$31.27}           & \textbf{29.31$\pm$52.37}         &  & \textbf{93.53$\pm$5.96}              & \textbf{85.72$\pm$11.11 }            & \textbf{5.834$\pm$5.13}        &  & \textbf{14.97$\pm$28.15}           &  & \textbf{96.62$\pm$3.06 }             & \textbf{89.03$\pm$8.32}              & \textbf{3.521$\pm$3.17}        \\ \specialrule{0.5mm}{1pt}{1pt}
		\end{tabular}
	}
	\end{table*}

	\section{Experiments}
	\subsection{Datasets}
	We evaluate our proposed JOINEDTrans on three retinal fundus image datasets for OD/OC segmentation and fovea detection: GAMMA, REFUGE, and PALM. On each dataset, we compare our method with representative SOTA methods and five-fold cross-validation is adopted for fair comparisons. 
	\subsubsection{GAMMA} The GAMMA\footnote{https://gamma.grand-challenge.org/} dataset \cite{R4} is provided by the GAMMA challenge organizers in the MICCAI2021 OMIA8 workshop. This dataset contains 200 fundus image data. The GAMMA challenge has three tasks: glaucoma classification, OD/OC segmentation, and fovea localization. The images are collected from multiple types of equipments, inducing diverse image resolutions that range from 1956 $\times$ 1934 to 2992 $\times$ 2000. We train our model on 100 training images and the validation split ratio is 0.2.
	
	\begin{table*}[!t]
		\centering
		\caption{Performance comparisons between our proposed JOINEDTrans and other SOTA methods, as evaluated on the REFUGE dataset. The best results are highlighted in bold, and the second-best results are underlined. }
		\label{tab:results-in-Refuge}
		\resizebox{\textwidth}{!}
		{%
			\begin{tabular}{lcccccccccccc}
				\specialrule{0.5mm}{1pt}{1pt}
				\multicolumn{13}{c}{REFUGE}                                                                                                                                                                                                                             \\  \specialrule{0.05mm}{1pt}{1pt}
				\multicolumn{1}{c}{\multirow{3}{*}{Method}} & \multicolumn{6}{c}{Coarse}                                                                                             &  & \multicolumn{5}{c}{Fine}                                                                         \\ \cline{2-7} \cline{9-13} 
				\specialrule{0em}{1pt}{1pt}
				\multicolumn{1}{c}{} & \multicolumn{2}{c}{Detection}               &  & \multicolumn{3}{c}{Segmentation}                                      &  & Detection             &  & \multicolumn{3}{c}{Segmentation}                                      \\  
				\multicolumn{1}{c}{} & Fovea $AED$ $\downarrow$ & OD $AED$ $\downarrow$  &           & OD $Dice$ (\%) $\uparrow$ & OC $Dice$ (\%) $\uparrow$ & $vCDR$ (\%) $\downarrow$   &           & Fovea $AED$ $\downarrow$ &           & OD $Dice$ (\%) $\uparrow$ & OC $Dice$ (\%) $\uparrow$ & $vCDR$ (\%) $\downarrow$   \\ \specialrule{0.05mm}{1pt}{1pt}
				DeepLabv3+  \cite{R40}          & -                     & -                    &           & 80.16$\pm$17.57             & 70.18$\pm$26.19             & 19.27$\pm$15.90         &           & -                     &           & 85.66$\pm$14.27             & 71.71$\pm$28.07             & 16.27$\pm$12.37         \\
				UNet++  \cite{R29}          & -                     & -                    &           & 81.37$\pm$15.61             & 68.27$\pm$29.38             & 18.61$\pm$14.81         &           & -                     &           & 86.15$\pm$13.81             & 72.79$\pm$26.57             & 14.55$\pm$11.87         \\
				UNet \cite{R30}                  & -                     & -                    &           & 84.45$\pm$13.71             & 72.30$\pm$24.51             & 14.32$\pm$10.97         &           & -                     &           & 90.22$\pm$10.14             & 73.46$\pm$27.35             & 14.00$\pm$10.71         \\
				UNet (ResNet50)         & 105.10$\pm$127.57         & 132.45$\pm$141.24        &           & 83.57$\pm$13.17             & 78.48$\pm$23.13             & 16.27$\pm$12.76         &           & -                     &           & 90.03$\pm$6.91              & 78.05$\pm$22.91             & 15.45$\pm$11.39         \\
				UNet++ (ResNet50)       & 98.15$\pm$112.56          & 82.57$\pm$128.84         &           & 83.72$\pm$10.25             & 76.99$\pm$14.45             & 15.37$\pm$11.95         &           & -                     &           & 92.52$\pm$6.18              & 83.71$\pm$18.23             & 14.43$\pm$10.24         \\
				DeepLabv3+ (ResNet101)  & 79.27$\pm$84.57           & 81.54$\pm$98.15          &           & 84.18$\pm$11.34             & 77.62$\pm$23.97             & 13.64$\pm$12.38         &           & -                     &           & 94.97$\pm$5.38              & 83.91$\pm$10.15             & 12.75$\pm$9.47          \\ \specialrule{0.05mm}{1pt}{1pt}
				Pixel-Wise Regression \cite{R57}  & 50.42$\pm$68.22           & 34.75$\pm$52.10          &           & -                       & -                       & -                   &           & 42.18$\pm$57.27           &           & -                       & -                       & -                   \\
				H-DenseUNet  \cite{R56}      & -                     & -                    &           & 85.17$\pm$10.73             & 76.51$\pm$20.60             & 16.71$\pm$15.09         &           & -                     &           & 91.02$\pm$7.21              & 80.16$\pm$19.12             & 15.99$\pm$11.26         \\
				Attention UNet \cite{R10}      & -                     & -                    &           & 84.36$\pm$9.71              & 80.16$\pm$15.73             & 11.25$\pm$10.47         &           & -                     &           & 94.35$\pm$7.35              & 82.84$\pm$13.27             & 12.58$\pm$9.33          \\
				JOINED   \cite{R13}       & \renewcommand{\ULdepth}{0.7pt}\uline{40.21$\pm$51.12}           & \renewcommand{\ULdepth}{0.7pt}\uline{29.53$\pm$35.19}          &           & 86.15$\pm$11.25             & 80.37$\pm$12.60             & 8.648$\pm$8.33          &           & 30.40$\pm$36.71           &           & 95.35$\pm$6.12              & 86.94$\pm$8.84              & \renewcommand{\ULdepth}{0.7pt}\uline{3.831$\pm$2.05}          \\ \specialrule{0.05mm}{1pt}{1pt}
				Segtran  \cite{R28}               & -                     & -                    &           & 90.15$\pm$5.27              & 85.29$\pm$13.58             & 5.319$\pm$4.52          &           & -                     &           & \renewcommand{\ULdepth}{0.7pt}\uline{96.08$\pm$4.25}              & 87.22$\pm$8.11              & 4.129$\pm$3.14          \\
				TransUnet \cite{R19}               & -                     & -                    &           & 94.68$\pm$2.70              & 85.87$\pm$7.68              & 5.343$\pm$4.64          &           & -                     &           & 95.27$\pm$4.30              & \renewcommand{\ULdepth}{0.7pt}\uline{88.66$\pm$7.39}              & 4.345$\pm$4.18          \\
				Swin-Unet \cite{R18}               & -                     & -                    &           & \renewcommand{\ULdepth}{0.7pt}\uline{94.68$\pm$2.63}              & \renewcommand{\ULdepth}{0.7pt}\uline{86.86$\pm$6.78}              & \renewcommand{\ULdepth}{0.7pt}\uline{4.658$\pm$3.87}          &           & -                     &           & 95.12$\pm$3.83              & 87.67$\pm$7.74              & 4.563$\pm$4.19          \\
				Proposed                & \textbf{35.27$\pm$42.29}  & \textbf{25.57$\pm$29.31} & \textbf{} & \textbf{95.54$\pm$1.93}     & \textbf{89.02$\pm$6.10}     & \textbf{4.067$\pm$3.58} & \textbf{} & \textbf{27.15$\pm$20.28}  & \textbf{} & \textbf{96.24$\pm$4.02}     & \textbf{90.18$\pm$7.57}     & \textbf{3.764$\pm$4.05} \\ \specialrule{0.5mm}{1pt}{1pt}
			\end{tabular}
		}
	\end{table*}
	\subsubsection{REFUGE} This dataset is provided by REFUGE\footnote{https://refuge.grand-challenge.org/Home2020/} \cite{R43}, as part of MICCAI2019. There are a total of 400 images for training, 400 for validation, and 400 for testing. The resolution for the training data is 2124 $\times$ 2056 and that for the validation and testing data is 1634 $\times$ 1634.	
	
	\subsubsection{PALM} 
	The pathologic myopia (PALM)\footnote{https://aistudio.baidu.com/aistudio/competition/detail/86/0/introduction} dataset is provided by the ISBI2019 Pathologic Myopia Ophthalmology Competition organizers. It contains 800 training images and 400 testing images. The image resolution is either 1444 $\times$ 1444 or 2124 $\times$ 2056. 
	There is no ground truth segmentation for OC, and thus there is no OC evaluation on this dataset. 
	For PALM, our validation split ratio is also 0.2.
	
	\subsection{Implementation Details}
	The proposed JOINEDTrans pipeline is implemented with Pytorch, using NVIDIA GeForce RTX 2080Ti GPUs. We use Swin-Transformer \cite{R23} as the encoder for all the three JSDM, FSM, and FLM modules. We employ the SGD optimizer with the following learning rate schedule \cite{R58}, 
	\begin{equation}\label{eq:learning_rate}
	L_r  = L_0\times\left(1.0 - \frac{iter\_num}{max\_iterations}\right)^{0.9} ,
	\end{equation}
	where $L_0$ is the initial learning rate which is set to be 0.05. In our experiments, we set 
	the trade-off coefficients $\lambda_0$, $\lambda_1$, and $\sigma$ to be respective 1, 1, and $h/100$. The hyper-parameters $\mu$ and $\nu$ for balancing the star-shape loss and the $Dice$ loss are both set to be 0.5. The Euclidean distance threshold $d_{FLM}$ in FLM is set to be 30. 
	The training time is about 10 hours for 300 epochs on GAMMA and 24 hours for 300 epochs on both PALM and REFUGE. The test time is about 0.5 seconds for a 2992 $\times$ 2000 image.
	\par We identify the smallest rectangle that contains the entire field of view and use the identified rectangle to crop each fundus image. We then resize all cropped images to 224 $\times$ 224 before being inputted to the network. The batch-size and window-size of the network are respectively set to be 12 and 7. The augmentation strategy we employ in training JOINEDTrans is as follows. The color distortion operation adjusts the brightness, contrast, and saturation of the images with a random factor in [-0.1, 0.1]. 
	Horizontal and vertical flipping as well as rotation operations are applied with a probability of 0.5 and Gamma noise is applied with a random factor in [-0.2, 0.2]. For the resizing operation, we randomly sample in [1/1.1, 1.1] and then times the original size. 
	For cropping outputs from the coarse stage, we empirically identify 448 $\times$ 448 to be an optimal size for OD/OC segmentation and 128 $\times$ 128 for fovea localization.

	
	
	\begin{table}[]
		\setlength\tabcolsep{3pt}
		\centering
		\caption{Performance comparisons between our proposed JOINEDTrans and other SOTA methods on the PALM dataset. The best results are highlighted in bold, and the second-best results are underlined. }
		\label{tab:results-in-PALM}
		\resizebox{\linewidth}{!}
		{\begin{tabular}{lcccccc}
				\specialrule{0.5mm}{1pt}{1pt}
				\multicolumn{7}{c}{PALM}                                                                                                                                                                        \\ 
				\specialrule{0.05mm}{1pt}{1pt}  
				\multirow{3}{*}{Method} & \multicolumn{3}{c}{Coarse}                                                                   &  & \multicolumn{2}{c}{Fine}                                            \\ \cline{2-4}  \cline{6-7}  \specialrule{0em}{1pt}{1pt}
				& \multicolumn{2}{c}{Detection}                                   &  Segmentation            &  & Detection             &  Segmentation                             \\ 
				& Fovea $AED$ $\downarrow$                     & OD $AED$     $\downarrow$ & OD $Dice$ (\%) $\uparrow$ &  & Fovea $AED$ $\downarrow$ & OD $Dice$ (\%) $\uparrow$                  \\ \specialrule{0.05mm}{1pt}{1pt}
				DeepLabv3+ \cite{R40}             & -                                         & -                   & 53.29$\pm$44.11         &  & -                     & 65.53$\pm$35.18                          \\
				UNet++ \cite{R29}                 & -                                         & -                   &  60.87$\pm$32.54         &  & -                     &  76.82$\pm$22.66                          \\
				UNet \cite{R30}                   & -                                         & -                   &  68.61$\pm$29.00         &  & -                     &  80.59$\pm$20.99                          \\
				UNet (ResNet50)         & 156.13$\pm$243.52                         & 156.37$\pm$399.47   &  54.33$\pm$40.12         &  & 90.51$\pm$60.70       &  69.75$\pm$32.67                          \\
				UNet++ (ResNet50)       & 140.52$\pm$271.15                         & 114.08$\pm$307.40   &  62.11$\pm$33.75         &  & 105.00$\pm$77.81      &  82.64$\pm$23.12                          \\
				DeepLabv3+ (ResNet101)  & 108.35$\pm$125.35                         & 92.58$\pm$182.67    &  75.12$\pm$23.74         &  & 80.21$\pm$52.70       &  92.79$\pm$7.76                           \\ \specialrule{0.05mm}{1pt}{1pt}
				Pixel-Wise Regression \cite{R57}   & 64.57$\pm$51.37                           & 53.72$\pm$68.28     &  -                       &  & 51.59$\pm$75.98       &  -                                        \\
				H-DenseUNet \cite{R56}            & -                                         & -                   &  69.59$\pm$35.92         &  & -                     &  85.33$\pm$12.51                          \\
				Attention UNet \cite{R10}         & -                                         & -                   &  70.94$\pm$24.21         &  & -                     &  87.76$\pm$9.51                           \\
				JOINED  \cite{R13}                & \renewcommand{\ULdepth}{0.7pt}\uline{53.47$\pm$43.02}                           & \renewcommand{\ULdepth}{0.7pt}\uline{38.28$\pm$46.25}     &  82.64$\pm$23.12         &  & \renewcommand{\ULdepth}{0.7pt}\uline{40.15$\pm$33.75}       &  \renewcommand{\ULdepth}{0.7pt}\uline{94.53$\pm$6.51}                           \\ \specialrule{0.05mm}{1pt}{1pt}
				Segtran \cite{R28}                & -                                         & -                   &  85.15$\pm$17.21         &  & -                     &  94.34$\pm$4.98                           \\
				TransUnet  \cite{R19}             & -                                         & -                   &  \renewcommand{\ULdepth}{0.7pt}\uline{94.08$\pm$10.18}         &  & -                     &  93.84$\pm$9.84                           \\
				Swin-Unet  \cite{R18}              & -                                         & -                   &  93.40$\pm$8.02          &  & -                     &  93.95$\pm$8.07                           \\
				Proposed                & \textbf{48.42$\pm$40.81} & \textbf{27.73$\pm$26.05}     &  \textbf{94.81$\pm$8.48}          &  & \textbf{35.22$\pm$28.50}       &  \textbf{95.95$\pm$4.47} \\ \specialrule{0.5mm}{1pt}{1pt}
			\end{tabular}
			
		}
	\end{table}

	\subsection{Evaluation Indicators}
	To assess the quality of the segmentation and detection results, five evaluation metrics including the Average Euclidean Distance ($AED$, pixel) in terms of both OD center and fovea, $Dice$ ($\%$) of both OD and OC, Mean Absolute Error ($MAE$) in $vCDR$ ($\%$) are employed. Among them, $AED$ evaluates the performance on the detection task; a small $AED$ is desirable. Both $Dice$ and $MAE$ evaluate the performance on the segmentation task; a large $Dice$ and a small $vCDR$ are desirable. 
	\subsection{Comparison to SOTA}	
	In Tables \ref{tab:results-gamma}$-$\ref{tab:results-in-PALM},  we compare JOINEDTrans against representative SOTA methods, including three general models that are widely used in various segmentation tasks, namely UNet \cite{R30}, UNet++ \cite{R29} and  DeeplabV3+ \cite{R40}. 
	Since these three models originally only output segmentation results, we add a branch that outputs coordinates, so that each general model can output both segmentation and detection results. We further replace the encoder of each of the three general methods with ResNet \cite{R72} to increase the amount of parameters for fairer comparisons with other SOTA methods. 
	\par
\begin{figure}
		\centering
		\includegraphics[width=\linewidth]{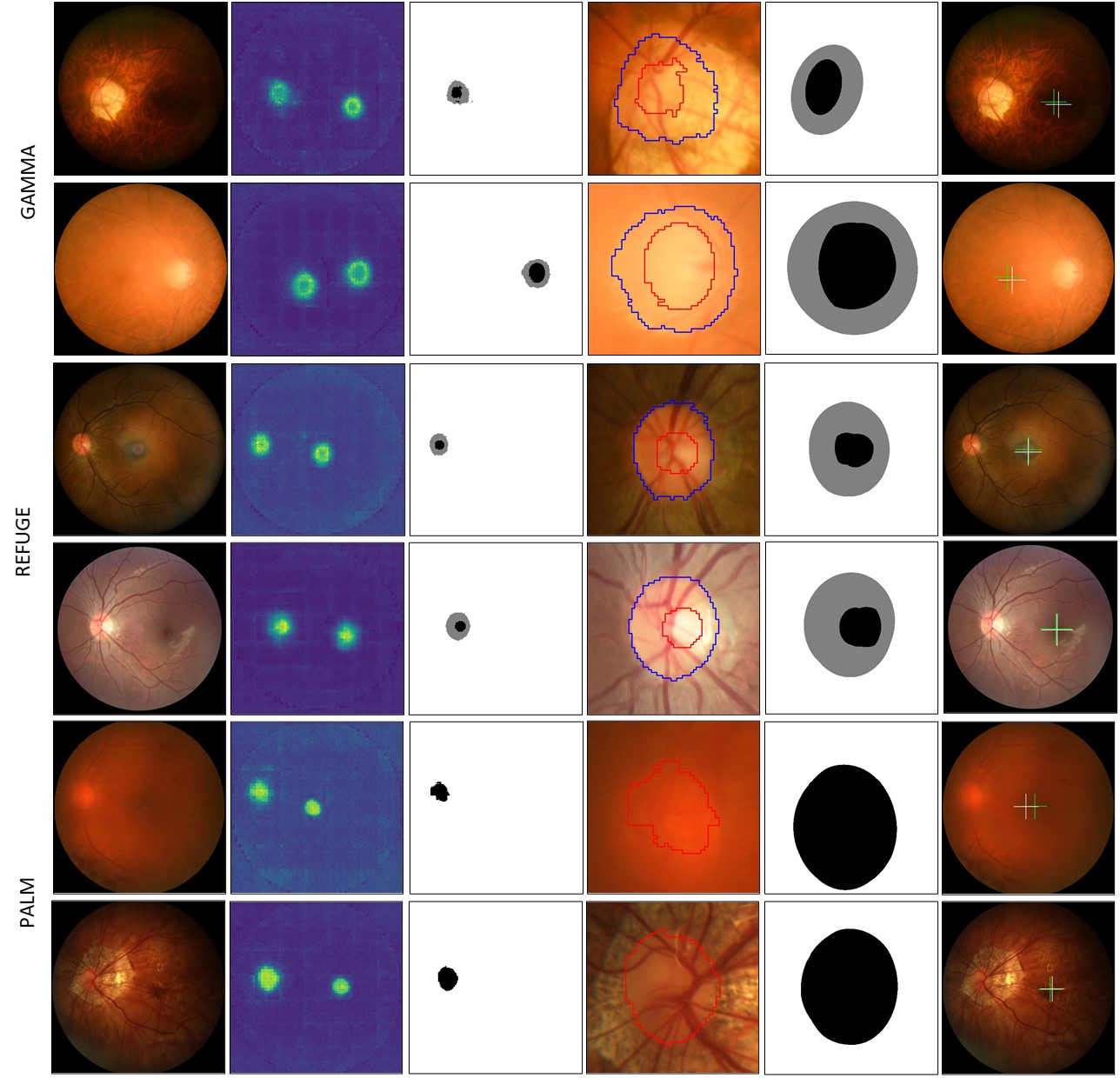}
		\caption{ Visualization of the JSDM outputs obtained on representative images from the three datasets. From left to right,  retinal fundus images, heatmaps obtained from the detection branch, coarse segmentation results obtained from the segmentation branch, cropped coarse segmentation results, ground truth segmentation results, coarse localization results wherein green is the ground truth and white is the prediction. From top to bottom are representative cases from GAMMA, REFUGE, and PALM.}
		\label{fig:result_all}
	\end{figure}

\begin{figure}[]
		\centering
		\includegraphics[width=\linewidth]{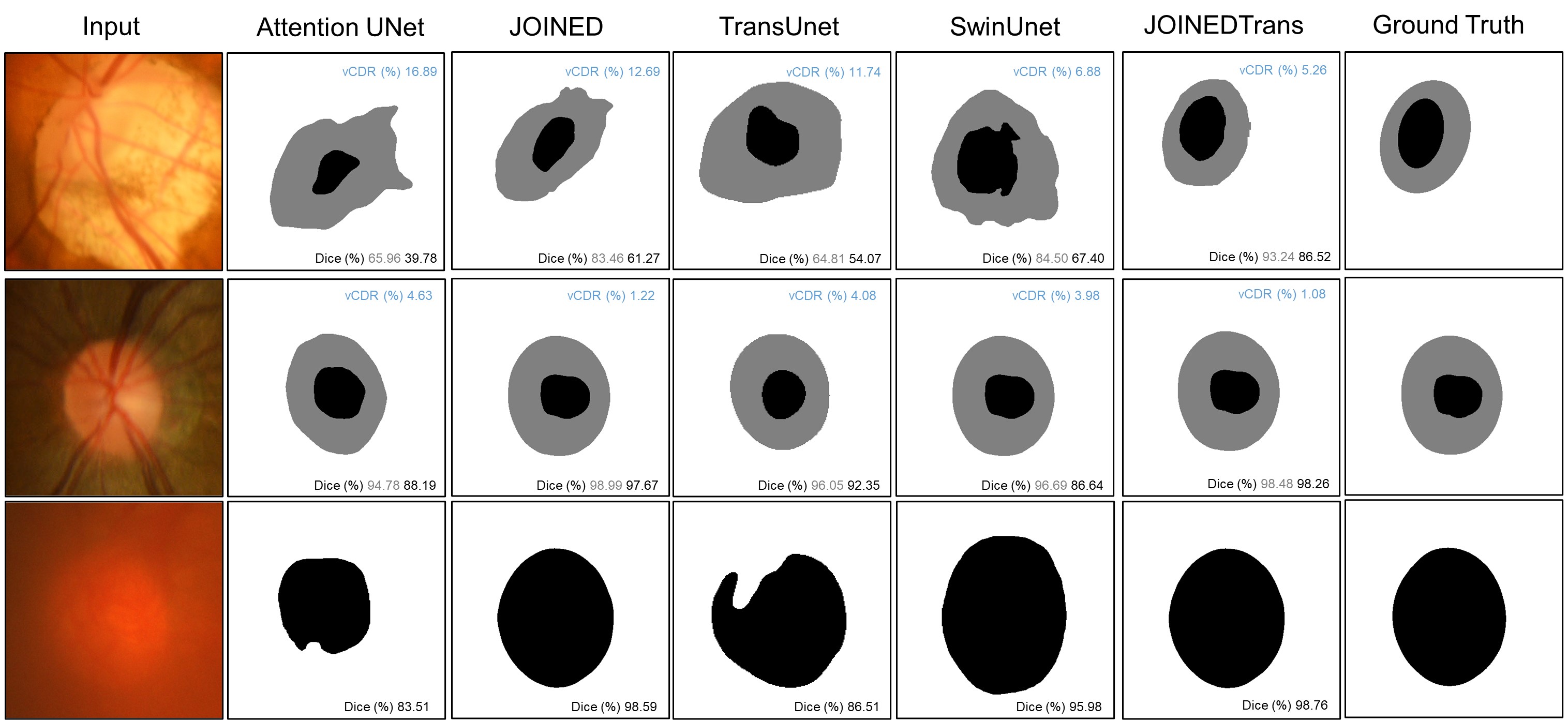}
		\caption{ Representative segmentation results from different automated methods and the ground truth. From top to bottom are representative cases from GAMMA, REFUGE, and PALM. For the GAMMA and REFUGE data, the gray area represents OD and the gray number represents the corresponding $Dice$ score of OD, the black area represents OC and the black number represents the corresponding $Dice$ score of OC, the blue number represents the corresponding $vCDR$ score. For the PALM data, the black area represents OD and the black number represents the corresponding $Dice$ score of OD.}
		\label{fig:compare}
	\end{figure}

 Apparently, our proposed JOINEDTrans outperforms all those three general segmentation models
	by very large margins. In those three tables, we also compare JOINEDTrans with seven recently-developed CNN-based and transformer-based methods, including Pixel-Wise Regression \cite{R57}, H-DenseUNet \cite{R56}, Attention UNet \cite{R10}, JOINED \cite{R13}, Segtran \cite{R28}, TransUnet \cite{R19}, and Swin-Unet \cite{R18}. 
	For fair comparisons, we conduct experiments for both coarse and fine stages for all methods. In the coarse stage, we find that the segmentation and detection results of JOINEDTrans are superior to all other methods, and its coarse segmentation even outperforms some other method's fine segmentation. 
	It is worth noting that JOINEDTrans is much better than other methods when assessed by the $vCDR$ metric which is a very critical index for clinical diagnoses of glaucoma \cite{R53}. 
	Fig. \ref{fig:result_all} shows representative visualization results obtained from JOINEDTrans's detection and segmentation branches. The representative examples selected from GAMMA and PALM are challenging samples with strong interference, while those from REFUGE are relatively simple samples with no interference. 
	Comparing the outputs of the detection and segmentation branches, it can be observed that the shape of the heatmap region of OD is highly correlated with its segmentation result. The reason may be that the detection branch guides the segmentation branch to some extent. Comparing the coarse segmentation with the ground truth, our proposed method not only performs comparably to manual delineation on simple samples, but also shows good performance on difficult ones. Similar conclusions can be also drawn by observing the coarse detection results.
	Representative OD/OC fine segmentation results on the three datasets are shown in Fig. \ref{fig:compare}. Due to the addition of the star-shape loss based regularization term in JOINEDTrans, the segmented edges are smoother compared to other methods. Clearly, the segmentation results of both OD and OC produced by JOINEDTrans are more precise and more accurate than those produced by other compared methods, in terms of both $Dice$ and $vCDR$ scores.

	\subsection{Ablation Study}
	
	Our proposed multi-task learning framework in the coarse stage is composed of three components, including a vessel pretrained block, a detection branch, and a segmentation branch. To verify the contribution of each of them, we construct five variants of the coarse stage of JOINEDTrans with the same fine stage and conduct ablation studies on the aforementioned three datasets. The ablation analysis results are tabulated in Table \ref{tab:ablation-all}. Model \uppercase\expandafter{\romannumeral1} and model \uppercase\expandafter{\romannumeral2} are respectively the baseline segmentation network and the baseline detection network. Model \uppercase\expandafter{\romannumeral3} consists of both the detection branch and the segmentation branch. It shows that when these two tasks are performed together, the performance of both fovea localization and OD/OC segmentation gets improved.  
	Model \uppercase\expandafter{\romannumeral4} and model \uppercase\expandafter{\romannumeral5} show the benefits of the vessel segmentation pretrained encoder respectively exerted to the segmentation task and the detection task. Proposed$^\star$ represents that with all three components included but no incorporation of the star-shape loss. Finally, the best results are obtained when all three components as well as the star-shape loss are included (our proposed JOINEDTrans). Please note, the fine stage is included in all compared models in Table \ref{tab:ablation-all}. 
	\begin{table*}[t]
		\centering
		\caption{Coarse stage ablation analysis results on the three datasets. A B and C respectively represent GAMMA, REFUGE, and PALM. Proposed$^\star$ means no star-shape loss.}
		\label{tab:ablation-all}
		\vspace{-2mm}
		\resizebox{\textwidth}{!}{
			\begin{tabular}{@{}cccccccccccccccccccccccc@{}}
				\specialrule{0.5mm}{1pt}{2pt}
				\multirow{3}{*}{Methods}                     & \multicolumn{3}{c}{Component}  &  & \multicolumn{7}{c}{Detection} &  & \multicolumn{11}{c}{Segmentation} \\ \cmidrule(r){2-4} \cmidrule(lr){6-12} \cmidrule(l){14-24} \specialrule{0mm}{1pt}{0pt}
				\multicolumn{1}{c}{} &
				\multirow{2}{*}{VesPreBlock} & \multirow{2}{*}{HeatBlock} & \multirow{2}{*}{SegBlock} &  & \multicolumn{3}{c}{Fovea $AED$ $\downarrow$} &  & \multicolumn{3}{c}{OD $AED$ $\downarrow$} &  & \multicolumn{3}{c}{OD $Dice$ (\%) $\uparrow$} &  & \multicolumn{3}{c}{OC $Dice$ (\%) $\uparrow$} &  & \multicolumn{3}{c}{$vCDR$ (\%) $\downarrow$} \\
				&                              &                            &                           &  & A        & B        & C       &  & A       & B       & C      &  & A       & B       & C       &  & A         & B        & C    &  & A        & B       & C   \\
				\midrule
				\uppercase\expandafter{\romannumeral1} &                              &                            & $\checkmark$ &  & -        & -        & -       &  & -       & -       & -      &  & 84.91   & 94.68   & 93.4    &  & 73.88     & 86.86    & -    &  & 10.98    & 4.658   & -   \\
				\uppercase\expandafter{\romannumeral2} &                              & $\checkmark$ &                           &  & 35.37    & 37.15    & 61.27   &  & 34.50   & 30.61   & 35.27  &  & -       & -       & -       &  & -         & -        & -    &  & -        & -       & -   \\
				\uppercase\expandafter{\romannumeral3} &                              & $\checkmark$                          & $\checkmark$                         &  & 30.17    & 36.5     & 53.19   &  & 31.57   & 27.46   & 30.21  &  & 86.53   & 95.02   & 93.71   &  & 76.27     & 88.12    & -    &  & 10.00    & 4.510   & -   \\
				\uppercase\expandafter{\romannumeral4} & $\checkmark$                            &                            & $\checkmark$                         &  & -        & -        & -       &  & -       & -       & -      &  & 90.32   & 95.14   & 93.81   &  & 80.24     & 88.37    & -    &  & 8.410    & 4.283   & -   \\
				\uppercase\expandafter{\romannumeral5} & $\checkmark$                            & $\checkmark$                          &                           &  & 27.24    & 35.79    & 51.60   &  & 30.10   & 26.05   & 30.88  &  & -       & -       & -       &  & -         & -        & -    &  & -        & -       & -   \\
				Proposed$^\star$                & $\checkmark$                            & $\checkmark$                          & $\checkmark$                         &  & 25.73    & 36.12    & 50.71   &  & 29.43   & 25.90   & 27.81  &  & 92.16   & 94.16   & 93.64   &  & 83.75     & 88.42    & -    &  & 7.124    & 4.208   & -  \\
				Proposed                & $\checkmark$                            & $\checkmark$                          & $\checkmark$                         &  & \textbf{25.24}    & \textbf{35.27}    & \textbf{48.42}   &  & \textbf{29.31}   & \textbf{25.57}   & \textbf{27.73}  &  & \textbf{93.53}   & \textbf{95.54}   & \textbf{94.81}   &  & \textbf{85.72}     & \textbf{89.02}    & -    &  & \textbf{5.834}    & \textbf{4.067}   & -   \\ 	\specialrule{0.5mm}{1pt}{1pt}
			\end{tabular}%
		}
	\end{table*}

	\subsection{Input Resolution} 
\begin{figure}
		\centering
		\includegraphics[width=\linewidth]{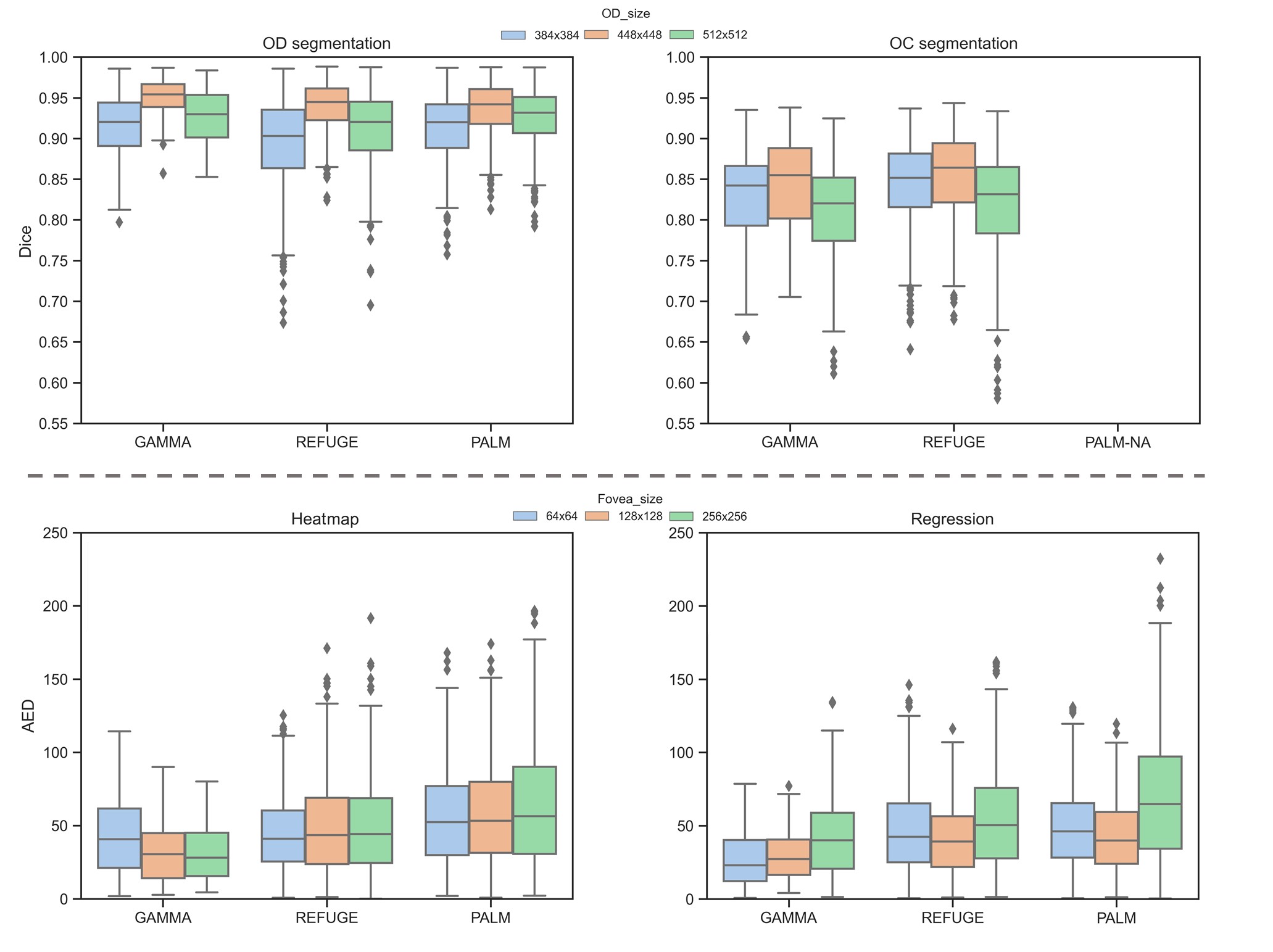}
		\caption{ Boxplots of the OD/OC segmentation and fovea localization performance with different input resolutions.}
		\label{fig:ablationbox}
	\end{figure}

	The resolution of the input image largely affects the performance of both segmentation and detection. For the GAMMA dataset, Fig. \ref{fig:ablationbox} shows that the OD/OC segmentation performance becomes worse when the resolution reduces from 448 $\times$ 448 to 384 $\times$ 384. A potential reason is that for some images the 384 $\times$ 384 resolution cannot fully cover OD, which highlights the importance of maintaining the structural integrity of OD/OC. Fig. \ref{fig:ablationbox} also demonstrates that the OD/OC segmentation performance becomes worse when the resolution increases from 448 $\times$ 448 to 512 $\times$ 512. This emphasizes that it is better to use a relatively small resolution while ensuring the integrity of OD/OC in FSM. Similar results are also observed on the other two datasets. As illustrated in the two bottom boxplots of Fig. \ref{fig:ablationbox}, in FLM, when the input resolution becomes smaller, the regression outputs become better but the heatmap outputs become worse. Balancing the performance of the two tasks, we choose 128 $\times$ 128 as the input resolution for FLM.

	\subsection{Cross-dataset Analysis}
	We also conduct cross-dataset analysis by training on GAMMA and testing on REFUGE, as well as vice versa, to analyze the generalizability and transferability of the learned representation in JOINEDTrans.
	As shown in Tables \ref{tab:results-in-REFUGEtoGAMMA} and \ref{tab:results-in-GAMMAtoREFUGE}, in both cross-dataset settings, our method has the smallest percentage of performance degradation compared to other methods. Looking at the data in Tables \ref{tab:results-gamma} and \ref{tab:results-in-REFUGEtoGAMMA}, we find that transformer-based methods drop less than CNN-based methods. The cross-dataset results are even better than within-dataset results for both Swin-Unet and Segtran. Nevertheless, as shown in Tables \ref{tab:results-in-Refuge} and \ref{tab:results-in-GAMMAtoREFUGE}, CNN-based methods are more stable than transformer-based ones. Considering that the REFUGE dataset has 800 images and the GAMMA dataset has only 100 images, we suspect that the amount of data has a great impact on transformer-based methods, as the performance of models trained on REFUGE does not drop significantly on GAMMA  whereas the models trained on GAMMA degrade largely when testing on REFUGE compared to within-dataset testing.
		\begin{table}[]
		\setlength\tabcolsep{3pt}
		\centering
		\caption{Performance comparisons between our proposed JOINEDTrans and other SOTA methods for training on REFUGE and prediction on GAMMA. The best results are highlighted in bold, and the second-best results are underlined. }
		\label{tab:results-in-REFUGEtoGAMMA}
		\resizebox{\linewidth}{!}
		{\begin{tabular}{lcccccc}
				\specialrule{0.5mm}{1pt}{1pt}
				\multicolumn{7}{c}{REFUGE to GAMMA}                                                                                                                                            \\ \specialrule{0.05mm}{1pt}{1pt}
				\multirow{2}{*}{Method}                           & 
				\multicolumn{2}{c}{Detection}                   &  & \multicolumn{3}{c}{Segmentation} \\
				& Fovea $AED$  $\downarrow$ & OD $AED$     $\downarrow$ &  & OD $Dice$ (\%) $\uparrow$ & OC $Dice$ (\%) $\uparrow$ & $vCDR$ (\%) $\downarrow$ \\ \specialrule{0.05mm}{1pt}{1pt}
				DeepLabv3+ \cite{R40}            & -                     & -                       &  & 63.16$\pm$17.57             & 61.18$\pm$26.19             & 24.27$\pm$21.90       \\
				UNet++ \cite{R29}                & -                     & -                       &  & 67.37$\pm$15.61             & 60.27$\pm$29.38             & 23.61$\pm$18.81       \\
				UNet \cite{R30}                  & -                     & -                       &  & 65.45$\pm$47.71             & 62.30$\pm$24.51             & 23.32$\pm$19.97       \\
				UNet (ResNet50)                                   & 105.10$\pm$127.57         & 132.45$\pm$141.24           &  & 68.57$\pm$13.17             & 68.48$\pm$23.13             & 20.27$\pm$18.76       \\
				UNet++ (ResNet50)                                 & 98.15$\pm$112.56          & 82.57$\pm$128.84            &  & 70.72$\pm$22.25             & 66.99$\pm$14.45             & 18.37$\pm$16.95       \\
				DeepLabv3+ (ResNet101)                            & 86.14$\pm$100.27          & 95.54$\pm$105.15            &  & 74.18$\pm$20.34             & 67.62$\pm$23.97             & 18.64$\pm$18.38       \\ \specialrule{0.05mm}{1pt}{1pt}
				Pixel-Wise Regression \cite{R57} & 60.32$\pm$78.44           & 45.07$\pm$61.08             &  & -                       & -                       & -                 \\
				H-DenseUNet \cite{R56}           & -                     & -                       &  & 78.17$\pm$19.73             & 71.21$\pm$20.60             & 16.71$\pm$15.09       \\
				Attention UNet \cite{R10}        & -                     & -                       &  & 76.21$\pm$17.71             & 70.16$\pm$21.73             & 16.25$\pm$18.47       \\
				JOINED  \cite{R13}               & \renewcommand{\ULdepth}{0.7pt}\uline{50.16$\pm$56.12}           & \renewcommand{\ULdepth}{0.7pt}\uline{40.18$\pm$49.51}             &  & 84.37$\pm$14.25             & 76.57$\pm$16.60             & \renewcommand{\ULdepth}{0.7pt}\uline{12.18$\pm$11.53}       \\ \specialrule{0.05mm}{1pt}{1pt}
				Segtran \cite{R28}               & -                     & -                       &  & 84.27$\pm$12.16             & 75.43$\pm$16.29             & 15.32$\pm$14.52       \\
				TransUnet  \cite{R19}            & -                     & -                       &  & 83.19$\pm$10.06             & 74.35$\pm$19.50             & 15.20$\pm$16.04      \\
				Swin-Unet  \cite{R18}             & -                     & -                       &  & \renewcommand{\ULdepth}{0.7pt}\uline{85.17$\pm$10.94}             & \renewcommand{\ULdepth}{0.7pt}\uline{76.70$\pm$16.21}             & 15.14$\pm$14.71       \\
				Proposed                                          & \textbf{40.12$\pm$50.18}           & \textbf{26.57$\pm$30.51 }            &  & \textbf{90.21$\pm$11.76 }            & \textbf{80.5+$\pm$14.95}             & \textbf{9.351$\pm$10.18}      \\ \specialrule{0.5mm}{1pt}{1pt}
			\end{tabular}
		}
	\end{table}
	
	\begin{table}[]
	
	\setlength\tabcolsep{3pt}
	\centering
	\caption{Performance comparisons between our proposed JOINEDTrans and other SOTA methods for training on GAMMA and prediction on REFUGE. The best results are highlighted in bold, and the second-best results are underlined.}
	\label{tab:results-in-GAMMAtoREFUGE}
	\resizebox{\linewidth}{!}
	{\begin{tabular}{lcccccc}
			\specialrule{0.5mm}{1pt}{1pt}
			\multicolumn{7}{c}{GAMMA to REFUGE}                                                                                                                                            \\ \specialrule{0.05mm}{1pt}{1pt}
			\multirow{2}{*}{Method}                           & 
			\multicolumn{2}{c}{Detection}                   &  & \multicolumn{3}{c}{Segmentation} \\
			& Fovea $AED$ $\downarrow$ & OD $AED$     $\downarrow$ &  & OD $Dice$ (\%) $\uparrow$ & OC $Dice$ (\%) $\uparrow$ & $vCDR$ $\downarrow$ \\ \specialrule{0.05mm}{1pt}{1pt}
			DeepLabv3+ \cite{R40}             & -                     & -                       &  & 67.24$\pm$21.57             & 64.27$\pm$25.19             & 20.13$\pm$17.90       \\
			UNet++ \cite{R29}                                                              & -                     & -                       &  & 70.19$\pm$19.61             & 65.04$\pm$25.38             & 19.57$\pm$16.81       \\
			UNet \cite{R30}                                                                & -                     & -                       &  & 74.16$\pm$15.71             & 68.40$\pm$23.51             & 19.27$\pm$18.97       \\
			UNet (ResNet50)                                                                                 & 124.16$\pm$129.57         & 145.97$\pm$147.24           &  & 73.26$\pm$17.17             & 72.48$\pm$19.13             & 18.72$\pm$16.76       \\
			UNet++ (ResNet50)                                                                               & 104.97$\pm$124.56         & 94.15$\pm$134.84            &  & 73.25$\pm$16.25             & 72.48$\pm$17.45             & 18.72$\pm$18.95       \\
			DeepLabv3+ (ResNet101)                                                                          & 81.67$\pm$87.57           & 88.19$\pm$107.15            &  & 74.61$\pm$18.34             & 72.63$\pm$23.97             & 17.38$\pm$16.38       \\ \specialrule{0.05mm}{1pt}{1pt}
			Pixel-Wise Regression \cite{R57}                                               & 62.27$\pm$75.22           & 39.10$\pm$71.10             &  & -                       & -                       & -                 \\
			H-DenseUNet \cite{R56}                                                         & -                     & -                       &  & 75.30$\pm$18.73             & 70.19$\pm$17.60             & 18.24$\pm$15.09       \\
			Attention UNet \cite{R10}                                                      & -                     & -                       &  & 79.34$\pm$15.71             & \renewcommand{\ULdepth}{0.7pt}\uline{75.34$\pm$15.73}             & 16.85$\pm$13.47       \\
			JOINED  \cite{R13}                                                             & \renewcommand{\ULdepth}{0.7pt}\uline{57.12$\pm$56.12}           & \renewcommand{\ULdepth}{0.7pt}\uline{35.17$\pm$57.19}             &  & \renewcommand{\ULdepth}{0.7pt}\uline{82.27$\pm$14.25}             & 74.60$\pm$16.60             & 15.65$\pm$14.33      \\ \specialrule{0.05mm}{1pt}{1pt}
			Segtran \cite{R28}                                                             & -                     & -                       &  & 82.17$\pm$16.27             & 70.85$\pm$13.58             & 15.39$\pm$14.52       \\
			TransUnet  \cite{R19}                                                          & -                     & -                       &  & 81.18$\pm$16.70             & 70.64$\pm$17.68             & 15.73$\pm$14.97       \\
			Swin-Unet  \cite{R18}                                                           & -                     & -                       &  & 80.17$\pm$15.94             & 70.74$\pm$16.21             & \renewcommand{\ULdepth}{0.7pt}\uline{15.14$\pm$16.71}       \\
			Proposed                                                                                        & \textbf{34.12$\pm$41.18}           & \textbf{31.59$\pm$51.51 }            &  & \textbf{83.77$\pm$10.76}             & \textbf{78.29$\pm$11.67}            & \textbf{11.41$\pm$9.96  }      	      \\ \specialrule{0.5mm}{1pt}{1pt}
		\end{tabular}
	}
\end{table}

	\section{Conclusion}
	In this paper, we explore the advantages of medical prior and multi-task learning, and pre-train an encoder for vessel segmentation based on domain generalization to help extract the spatial features of vessels. Taking advantage of transformer's ability to better extract global features, we propose and validate JOINEDTrans, a novel prior-guided multi-task transformer framework for joint OD/OC segmentation and fovea detection.
	Extensive experiments on three publicly accessible retinal fundus datasets indicate that our proposed method significantly outperforms SOTA methods on both segmentation and detection tasks, showing the effectiveness of the vessel prior and the joint learning strategy.
\bibliographystyle{IEEEtran}

\bibliography{joined}

\end{document}